# Intermediate anomalous Hall states induced by noncollinear spin structure in magnetic topological insulator $MnBi_2Te_4$

Jing-Zhi Fang,[1,2,3,*] Shuo Wang,[1,*] Xing-Guo Ye,[3,*] Ben-Chuan Lin,[1,#] An-Qi Wang,[3] Hao-Nan Cui,[1,3] Jian-Kun Wang,[1] Guang-Yu Zhu,[1] Song Liu,[1] Yongkai Li,[4,5] Zhiwei Wang,[4,5] Yugui Yao,[4,5] Zhongming Wei,[2,†] Dapeng Yu,[1] and Zhi-Min Liao[1,3,‡]

[1]*Shenzhen Institute for Quantum Science and Engineering, Southern University of Science and Technology, Shenzhen, 518055, China*

[2]*State Key Laboratory of Superlattices and Microstructures, Institute of Semiconductors, Chinese Academy of Sciences & Center of Materials Science and Optoelectronics Engineering, University of Chinese Academy of Sciences, Beijing 100083, China*

[3]*State Key Laboratory for Mesoscopic Physics and Frontiers Science Center for Nano-optoelectronics, School of Physics, Peking University, Beijing 100871, China*

[4]*Key Laboratory of Advanced Optoelectronic Quantum Architecture and Measurement, Ministry of Education, School of Physics, Beijing Institute of Technology, Beijing 100081, China*

[5]*Micronano Center, Beijing Key Lab of Nanophotonics and Ultrafine Optoelectronic Systems, Beijing Institute of Technology, Beijing 100081, China*

[*] *These authors contributed equally to this work.*

[#] linbc@sustech.edu.cn     [†] zmwei@semi.ac.cn    [‡] liaozm@pku.edu.cn

**Abstract:**

**The combination of topology and magnetism is attractive to produce exotic quantum matters, such as the quantum anomalous Hall state, axion insulators and the magnetic Weyl semimetals. $MnBi_2Te_4$, as an intrinsic magnetic topological insulator, provides a platform for the realization of various topological phases. Here we report the intermediate Hall steps in the magnetic hysteresis of $MnBi_2Te_4$, where four distinguishable magnetic memory states at zero magnetic field are revealed. The gate and temperature dependence of the magnetic intermediate states indicates the noncollinear spin structure in $MnBi_2Te_4$, which can be attributed to the Dzyaloshinskii-Moriya interaction as the coexistence of strong**

**spin-orbit coupling and local inversion symmetry breaking on the surface. Moreover, these multiple magnetic memory states can be programmatically switched among each other through applying designed pulses of magnetic field. Our results provide new insights of the influence of bulk topology on the magnetic states, and the multiple memory states should be promising for spintronic devices.**

The interplay between magnetism and electronic band topology in magnetic topological insulators provides an ideal platform to realize exotic topological states [1], such as quantum anomalous Hall states [2] and axion insulators [3,4]. In addition, magnetic topological insulators can also hold exotic spin texture in real space, such as Skyrmions [5]. The Dzyaloshinskii-Moriya interaction (DMI) would emerge in magnetic topological insulator with strong spin-orbit coupling (SOC) as breaking the inversion symmetry [6]. The DMI is generally crucial for creating and manipulating of real space spin textures [7-9], including the Skyrmions, promising in the field of spintronics [10].

Recently, an intrinsic magnetic topological insulator $MnBi_2Te_4$ has been theoretically proposed [3,11-13] and experimentally confirmed [14-21]. $MnBi_2Te_4$ is a van der Waals layered ternary tetradymite compound, consisting of stacking Te-Bi-Te-Mn-Te-Bi-Te septuple layers (SLs) along the $c$ axis [12]. Extensive efforts [11-32] have been focused on various topological phases in $MnBi_2Te_4$, including quantum anomalous Hall states [16], axion insulator states [12,13,17] and magnetic Weyl semimetals [12,13]. Moreover, the exotic spin structures have been revealed in $MnBi_2Te_4$. The magnetic ground state of $MnBi_2Te_4$ is proposed to be an A-type antiferromagnet [13]. The intralayer coupling leads to ferromagnetic (FM) order with out-of-plane easy axis in each SL; while neighboring SLs possess antiferromagnetic (AFM) exchange coupling [22-24]. Importantly, the so-called canted-AFM state, which is a noncollinear spin structure, can be formed by the spin-flop transition [30-32], inducing a net Berry curvature that contributes to the intrinsic anomalous Hall effect [18]. Additionally, strong spin fluctuations induced by spin-wave excitations [18] are also reported to exist in MnBi$_2$Te$_4$. These spin structures, together with strong SOC and bulk band topology in MnBi$_2$Te$_4$, are promising for exploring exotic spin quantum states [5].

Here, we report the surface noncollinear spin structure induced additional anomalous Hall resistance steps in the odd-number SL MnBi$_2$Te$_4$. Instead of the two already-known magnetic states in the magnetic hysteresis, four distinguishable magnetic states are identified, manifested by their different anomalous Hall resistance. All the four states could stably exist at zero external magnetic field, showing the nature of non-volatile memory. Furthermore, these multiple memory states can be switched among each other by applying designed magnetic field pulse. Systematical measurements as varying temperature and gate voltage suggest that the intermediate Hall states should be from a DMI induced surface noncollinear spin structure, consistent with the observed corresponding minimum longitudinal resistance with less spin related scattering.

The bulk crystal MnBi$_2$Te$_4$ was synthesized through flux methods [16], and the few-layer flakes were obtained by the mechanical exfoliation method. A typical 7-SLs MnBi$_2$Te$_4$ device with hall-bar electrodes and ~ 1 μm channel length (Supplemental Material, Fig. S1 [33]) was selected and measured. The measurement configuration is shown in the inset of Fig. 1(a), where the two-terminal resistance, four-terminal Hall resistance and longitudinal resistance are defined as $R_2 = \frac{V_2}{I}$, $R_{xy} = \frac{V_y}{I}$ and $R_{xx} = \frac{V_x}{I}$, respectively. A resistance kink was observed in the temperature dependence curve, corresponding to the Neel temperature ~21 K (Supplemental Material, Fig. S2 [33]). Figure 1(a) shows the transfer curve of the two-terminal resistance $R_2$ at 1 K for a typical 7-SLs MnBi$_2$Te$_4$ device (Device A). The charge neutral point with resistance maximum is around $V_g = 22$ V. Nearly-quantized plateau of $R_2$ was observed around the charge neutral point under magnetic field of 14 T (Figs. 1(a) and 1(b)). The two-terminal measurement under high magnetic field is a result of the difference of chemical potential of the two quantized edge states.

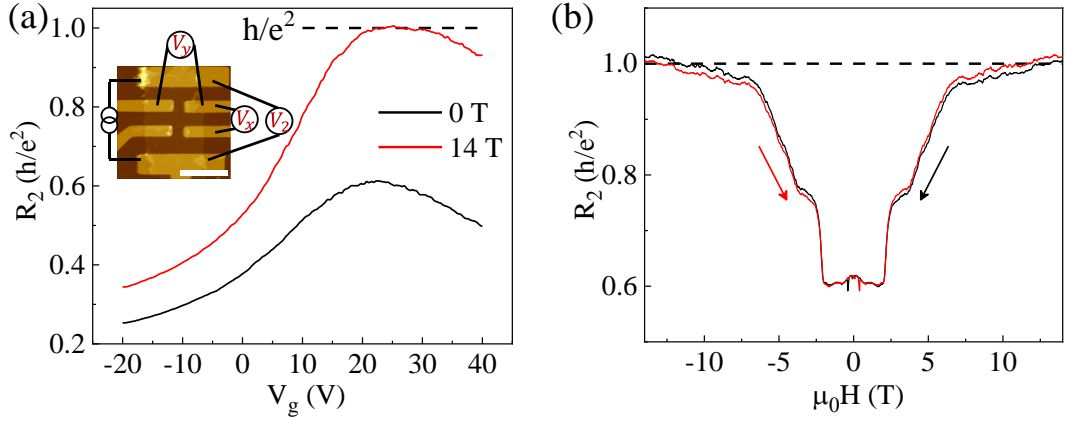

**FIG. 1. The two-terminal resistance $R_2$ measured at 1 K.** (a) The transfer curves under magnetic field of 0 and 14 T, respectively. Inset: the atomic force microscope image and the measurement configuration of a typical device. The scale bar is 4 μm. (b) The $R_2$ as a function of magnetic field near the charge neutral point $V_g = 22$ V.

In addition to the resistance plateau under high magnetic fields, the hysteresis loops of $R_{xy}$ are observed, as shown in Fig. 2(a). Intriguingly, multiple step-like changes are shown in the hysteresis loop, indicating multiple magnetic state transitions. Such multiple-step magnetic transitions are repeatedly observed under different gate voltages. Figure 2(b) shows the anomalous Hall resistance $R_{xy}^{AH}$ at 5 K and $V_g = 25$ V, obtained by subtracting the ordinary Hall resistance. The coercive fields $\mu_0 H_{c1}$ and $\mu_0 H_{c2}$ are defined at the first and second jumps of $R_{xy}^{AH}$, respectively, which are independent of the gate voltages, as shown in Fig. 2(a). Such gate independence is consistent with the localized character of the magnetic moment in MnBi$_2$Te$_4$ [12]. Then we also define $\Delta H_c$ as $\Delta H_c = H_{c2} - H_{c1}$, which also shows gate-independence. The four-step anomalous Hall resistances are defined as $R_1^{AH}$, $R_2^{AH}$, $R_3^{AH}$, $R_4^{AH}$, respectively, as shown in Fig. 2(b). The difference between the minor step and the major step is defined as $\Delta R^{AH} = R_4^{AH} - R_3^{AH}$, describing the intermediate step height. The anomalous Hall resistance shows non-monotonic gate dependence with a peak around the charge neutral point, as shown in Fig. 2(c) at 5 K (see Supplemental Material, Fig. S3 [33] at 10 K). This observation indicates that the anomalous Hall effect mainly originates from the intrinsic mechanism related to Berry curvature [34], rather than extrinsic effects such as side jump or skew scatterings.

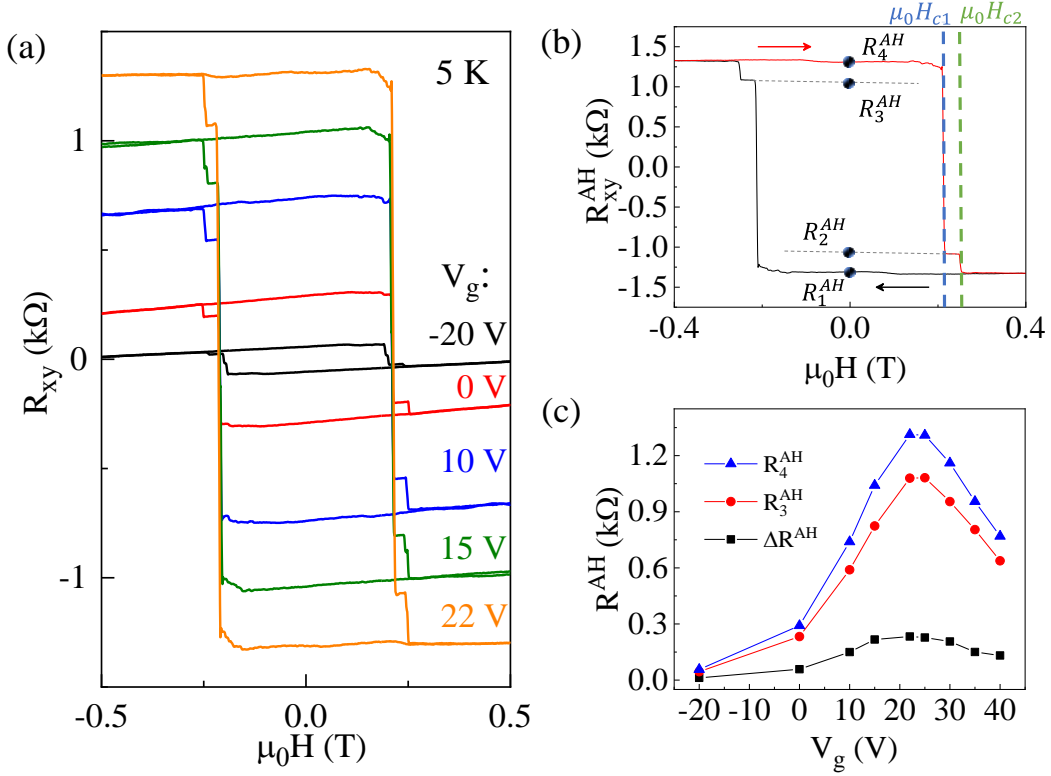

**FIG. 2. The gate-voltage dependence of Hall resistance at 5 K.** (a) The magnetic hysteresis at different gate voltages. (b) Anomalous Hall resistance $R_{xy}^{AH}$ at $V_g = 25$ V. The critical fields $H_{c1}$ and $H_{c2}$ corresponding to the two-step changes are denoted by the blue and green lines, respectively. The four anomalous Hall states $R_1^{AH}$, $R_2^{AH}$ and $R_3^{AH}$, and $R_4^{AH}$ are denoted by the blue balls. (c) The $R_3^{AH}$, $R_4^{AH}$ and the step height $\Delta R^{AH} = R_4^{AH} - R_3^{AH}$ as a function of gate voltage.

Figure 3(a) shows the temperature evolution of the magnetic hysteresis at $V_g = 25$ V. With increasing temperature beyond the Neel temperature (~21 K), the hysteresis gradually diminishes. Nevertheless, the intermediate Hall step is so robust that it persists as long as the magnetic hysteresis exists below the Neel temperature. The coercive fields and anomalous Hall resistance are decreased upon increasing temperature, as presented in Figs. 3(b) and 3(c), respectively. It is found that both the $\mu_0 \Delta H_c$ and $\Delta R^{AH}$ have a much weaker temperature dependence than that of the $\mu_0 H_{c1,c2}$ and $R_{3,4}^{AH}$, respectively, suggesting that the intermedia Hall step originates a mechanism differencing from the intrinsic magnetism of MnBi$_2$Te$_4$.

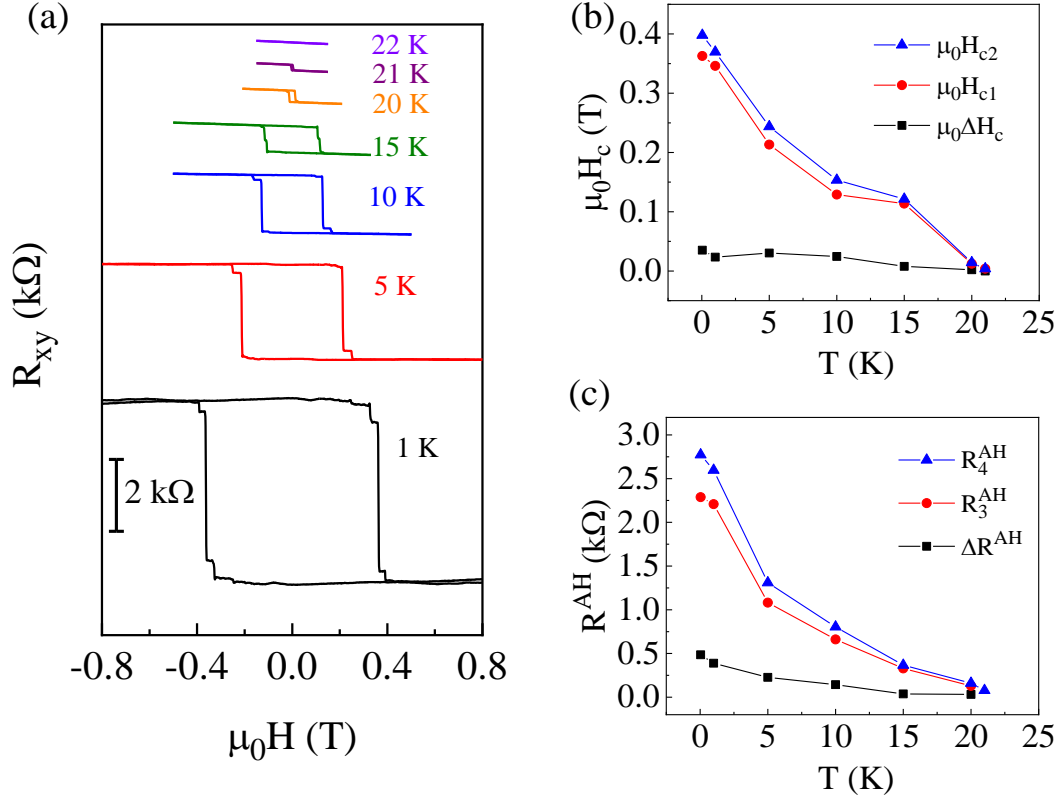

**FIG. 3. Temperature evolution of the anomalous Hall state.** (a) The temperature dependence of the hysteretic loop and the intermediate Hall step at $V_g = 25$ V. (b) The critical fields $H_{c1}$, $H_{c2}$ and $\Delta H_c$ at $V_g = 25$ V as a function of temperature. (c) The anomalous Hall resistance $R_4^{AH}$, $R_3^{AH}$ and step height $\Delta R^{AH}$ at $V_g = 25$ V as a function of temperature.

To further comprehend the emergent additional Hall step, minor loop measurements are performed at 1.5 K and $V_g = 25$ V. Minor loops 1 and 2 are shown in Figs. 4(a) and 4(b), respectively, where the magnetic field sweeping sequences are denoted by the arrows. The minor loop is robustly reproducible when repeating the sweeping cycles (Supplemental Material, Fig. S4 [33]). Taking the loop 1 as an example to illustrate the magnetic field sweeping details, which starts from a magnetic field larger than $\mu_0 H_{c2}$, then decreases to zero and further to the opposite direction, reaching to the intermediate Hall step, and then sweeps back. In contrast to the conventional hysteresis loop with two magnetic states, i.e., up or down states, here a total of four distinguishable magnetic memory states are realized at zero magnetic field, denoted as $|1\rangle$, $|2\rangle$, $|3\rangle$, $|4\rangle$ in Fig. 4(c).

As shown in Fig. 4(d), each of the four magnetic memory states can be uniquely obtained by applying well designed pulse of magnetic field. That is, the $|4\rangle$ and $|1\rangle$ states can be obtained by applying the negative and positive magnetic pulses with peak value larger than $H_{c2}$, respectively. The $|3\rangle$ and $|2\rangle$ states can be obtained by applying two successive magnetic pulses. Specifically, the two successive pulses with peak values $H_2$ and $-H_1$ produce the $|3\rangle$ state, and with peak values $-H_2$ and $H_1$ produces the $|2\rangle$ state, where the peak values satisfy $H_2 > H_{c2} > H_1 > H_{c1}$. The programmable four magnetic memory states presented in Fig. 4(d) are obtained from another 7-SLs MnBi$_2$Te$_4$ device (device B), giving the reproducibility from different devices. The device B also demonstrate the multiple memory states in the magnetic hysteresis loops of $R_{xy}$ (Supplemental Material, Fig. S5 [33]), reproducing the similar behaviors in device A.

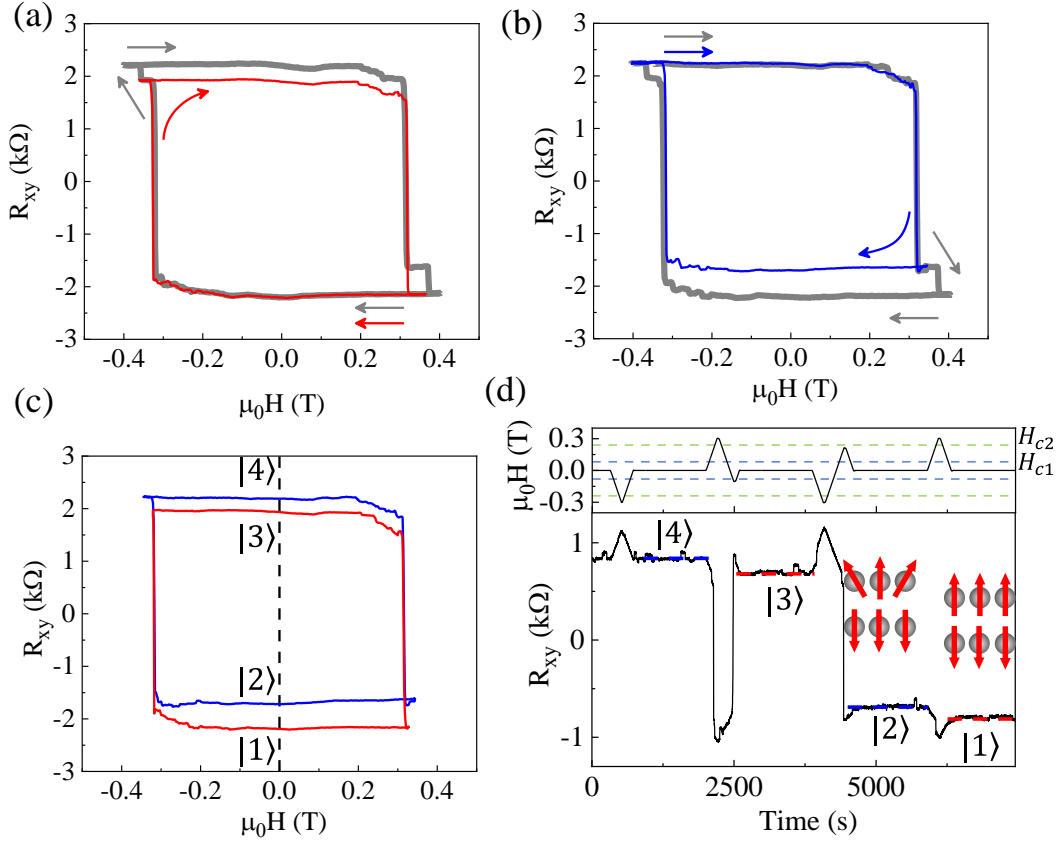

**FIG. 4. Minor loop measurements and programmable magnetic memory states.** (a) and (b) The minor loop measurements of device A at 1.5 K and $V_g = 25$ V correspond to different sweeping magnetic field sequence, as denoted by the arrows. The grey line represents the major loop, while the red (blue) line represents the minor loop 1(2). (c) The four magnetic memory states $|1\rangle$, $|2\rangle$, $|3\rangle$ and $|4\rangle$ at zero magnetic field are identified from two different sets of magnetic field sweeping hysteresis. (d) The programmable four magnetic memory states realized through the designed magnetic field pulses in device B at 15 K. Top panel: the designed pulses of magnetic fields. The dashed lines denote the coercive fields $\mu_0 H_{c1}$ and $\mu_0 H_{c2}$. Bottom panel: the four memory states achieved independently. Inset: The fully-aligned A-type AFM order and noncollinear spin texture for states $|1\rangle$ and $|2\rangle$, respectively. The gray spheres denote the Mn atoms in $MnBi_2Te_4$, while the red arrows denote the magnetic moments of these Mn atoms. Note here the magnetic order of only the top two SLs are illustrated.

Previously, the multiple-step behavior of hysteresis loop has been observed in FM/AFM superlattice [35], perovskite manganites [36], Cr-doped topological insulators [37], orbital Chern insulators [38]. These interesting intermediate magnetization steps are usually related to novel metamagnetism, which originates from different mechanisms in different systems. Here, the intermediate step states observed in 7-SLs MnBi$_2$Te$_4$ system are robust at zero field, inducing four distinct magnetic memory states, which may not originate from the multi-domain states due to the following reasons. First, such multi-domain states are generally not so robust, especially at zero magnetic field [39], in contrast to our results. Second, the domain configuration in MnBi$_2$Te$_4$ is reported to be randomized after a field cycle to the magnetic saturate state [31], while the intermediate Hall step behavior of our sample is robust against large field cycles up to 14 T (Supplemental Material, Fig. S6 [33]). Third, the anomalous Hall ratio $R_4^{AH}/\Delta R^{AH}$ is insensitive to temperature and gate voltages (Supplemental Material, Fig. S7 [33]), which may also indicate these intermediate states are induced by surface-related effect. Moreover, the magnetic domains should be randomly distributed for different samples [32] and the domain size is usually ~ 10 μm [31,32], while our results are well reproduced in different devices and the device channels are usually on the order of 1 μm.

Due to the A-type AFM structure, hysteresis loops can be observed under low magnetic field in the odd-SLs MnBi$_2$Te$_4$ with uncompensated magnetization. For the conventional magnetic hysteresis, two magnetic states can be identified at zero field, that is, the magnetization up or down states. In contrast, here a total of four magnetic memory states are observed [Fig. 4(c)]. Interestingly, for the additional intermediate states, the intermediate step height $\Delta R^{AH}$ demonstrates non-monotonic gate dependence with a peak around charge neutral point [Fig. 2(c)]. Such gate-tunability indicates the intrinsic Berry curvature-related origin of these additional intermediate states.

Here we attribute the intermediate Hall steps and the associated multiple memory states to the emergent surface noncollinear spin structure, which could contribute net Berry curvature. It is noted that a finite DMI term tends to induce noncollinear spin texture [5]. However, nonzero DMI requires strong SOC and the breaking of inversion symmetry [40]. The strong SOC is already present with nontrivial band topology. Although the inversion symmetry can persist in bulk MnBi$_2$Te$_4$ [12], the surface inversion symmetry can be naturally broken by the surface-related effect, such as surface atomic reconstruction, local defects, disorders and mechanical strains [29,41]. Thus, the surface spin magnetic moment may surfer a finite DMI, resulting in an intermediate magnetic state by the noncollinear spin structure. As denoted in Fig. 4(c), the $|1\rangle$ and $|4\rangle$ states, i.e., the lowest and topmost states, should be the fully-aligned states with perfect A-type AFM order. The $|2\rangle$ and $|3\rangle$ states, i.e., the intermediate states, hold noncollinear spin structure, where the spins deviate from the c-axis with nonzero in-plane component due to the DMI. Different from the perfect A-type AFM order with all spins along c-axis, the spins of the intermediate states between adjacent layer are no longer strictly antiparallel, thus named as noncollinear spin structure. Our explanation is consistent with the previous report, where the noncollinear spin structures appear in the AFM order with effective in-plane exchange bias [35]. In addition to the Hall resistance, the longitudinal resistance $R_{xx}$ is also carefully analyzed. It is found a plateau-like $R_{xx}$ dip occurs exactly at the intermediate Hall step (Supplemental Material, Fig. S8a [33]). This $R_{xx}$ dip is consistent with the noncollinear surface spin structure, which reduces the spin related scattering and decreases the longitudinal resistance. The spin related scattering is further evidenced by the $ln(T)$ dependence of the $R_{xx}$ dip (Supplemental Material, Fig. S8b [33]), indicating the dominated Kondo effect [42].

Interestingly, the surface-related effect is generally reported to exist in MnBi$_2$Te$_4$ [29,30,32], which is usually unavoidable during the material growth and device fabrication due to the susceptibility of MnBi$_2$Te$_4$ samples, being responsible for the emergent intermediate magnetic states. Moreover, similar intermediate magnetic states are also observed in hysteresis loop of a 7-SLs Sb-doped MnBi$_2$Te$_4$ [43]. It is worth noting that in a previously reported 5-SLs MnBi$_2$Te$_4$ sample which shows quantized plateau [16], such intermediate magnetic states are absent, which should be attributed to the extremely high quality of the sample that suppresses the surface-related effect. For our work, the Hall resistance at 1 K is not quantized but reaches a fraction of the quantized plateau ($\sim 0.1 h/e^2$) at zero field [see Fig. 3(a)]. The absence of perfect quantization should also be related to the surface-related effect. These surface-related effect could effectively reduce the out-of-plane magnetization, thus reducing the exchange gap, which should be responsible for the absence of perfect quantization.

In summary, our detailed magnetotransport results in MnBi$_2$Te$_4$ reveal multiple distinguishable magnetic memory states induced by surface noncollinear spin structure. The noncollinear spin texture in real space would lead to a net Berry curvature and thus result in the intermediate memory states. These memory states are reproducible and stable, and each of these states can be solely achieved by applying designed magnetic pulses. The emergence of the noncollinear spin structure could be attributed to the DMI between the surface spins, resulted by the combination of strong SOC and surface inversion symmetry breaking. Our findings suggest that the interplay between the topological electronic states and magnetic order can produce exotic spin textures based on strain-engineering [44] or symmetry-engineering [45], paving the way to manipulate the spin structures in numerous magnetic materials with strong SOC.

Acknowledgements: This work was supported by the Key-Area Research and Development Program of Guangdong Province (Grant No. 2020B0303060001, No. 2018B030327001, No. 2018B030326001), National Key Research and Development Program of China (No. 2020YFA0309300, No. 2018YFA0703703, and No. 2017YFA0207500), National Natural Science Foundation of China (No. 12004158, No. 12074162, No. 91964201, No. 61825401, and No. 11774004), the Strategic Priority Research Program of Chinese Academy of Sciences (Grant No. XDB43000000), the Natural Science Foundation of Guangdong Province (Grant No. 2017B030308003), and the Guangdong Provincial Key Laboratory (Grant No. 2019B121203002).

Liao, Strain Tunable Berry Curvature Dipole, Orbital Magnetization and Nonlinear Hall Effect in WSe$_2$ Monolayer, *Chin. Phys. Lett.* **38**, 017301 (2021).

[45] T. Akamatsu, T. Ideue, L. Zhou, Y. Dong, S. Kitamura, M. Yoshii, D. Yang, M. Onga, Y. Nakagawa, K. Watanabe *et al.*, A van der Waals interface that creates in-plane polarization and a spontaneous photovoltaic effect, *Science* **372**, 68 (2021).

# Supplemental Material

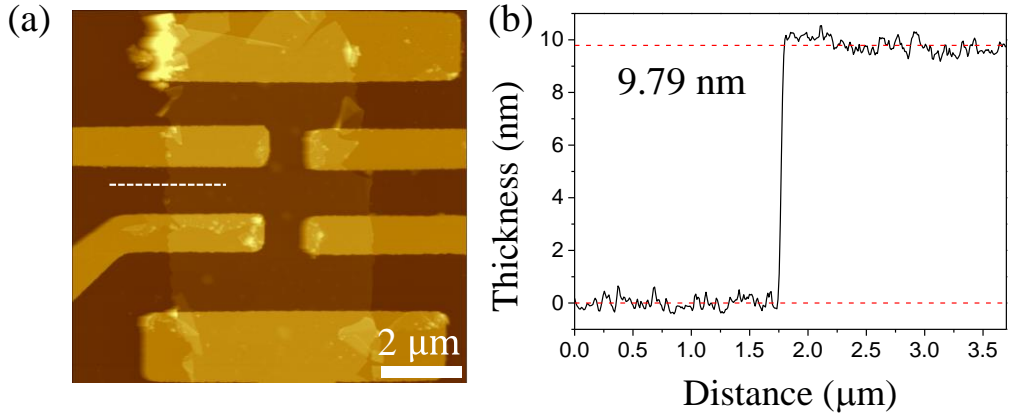

**FIG. S1. Characterizations of a typical MnBi$_2$Te$_4$ device.**

(a) The AFM image of the device.

(b) The corresponding white line profile of the device, revealing that the thickness is ~9.79 nm, which corresponds to 7-SL of MnBi$_2$Te$_4$.

The MnBi$_2$Te$_4$ thin flakes were first obtained by mechanical exfoliation method onto a SiO$_2$/Si substrate in an argon-filled glove box with O$_2$ and H$_2$O content below 0.01 parts per million to avoid sample degeneration. The Si substrate was covered with a 285 nm thick SiO$_2$, which was also used as back gate $V_g$. Before the whole device fabrication, the surrounding thick flakes were removed by a sharp needle. The Cr/Au electrodes were prepared by electron beam evaporation, after the hall-bar pattern was prepared by standard electron beam lithography. Most of the fabrication processes were done in the glove box. When the device was taken out of the glove box, the device was always covered by a layer of polymethyl methacrylate (PMMA) or h-BN to mitigate air contamination.

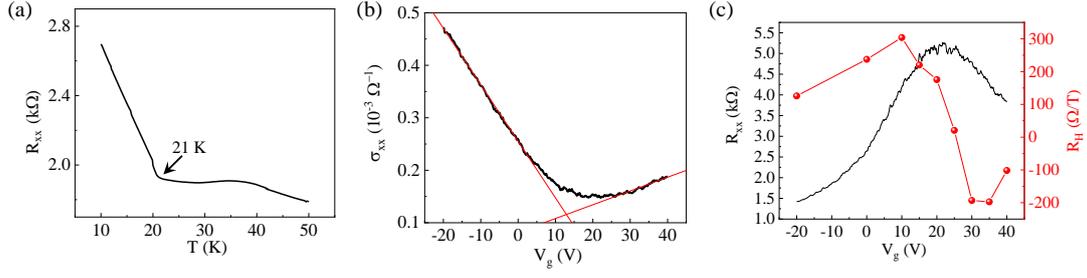

**FIG. S2. The basic properties of the MnBi$_2$Te$_4$ sample.**

(a) The temperature dependent longitudinal resistance indicates the Neel temperature is around 21 K.

(b) The four-terminal conductivity $\sigma_{xx}$ as a function of back gate $V_g$ at 1 K in Device A. The mobility is estimated by the formula $\mu = \frac{\partial \sigma_{xx}}{\partial V_g} \frac{t}{\varepsilon_0 \varepsilon_{ox}}$, where $t$ is the thickness of the oxide layer, $\varepsilon_0$ is vacuum dielectric constant and $\varepsilon_{ox}$ is the dielectric constant of SiO$_2$. The mobility is estimated to be around 900 cm$^2$/Vs for holes and 220 cm$^2$/Vs for electrons.

(c) Four-terminal longitudinal resistance $R_{xx}$ and Hall coefficient $R_H$ as a function of back gate $V_g$ in Device A at 1 K. Around the charge neutral point, $R_{xx}$ shows a peak while $R_H$ changes the sign.

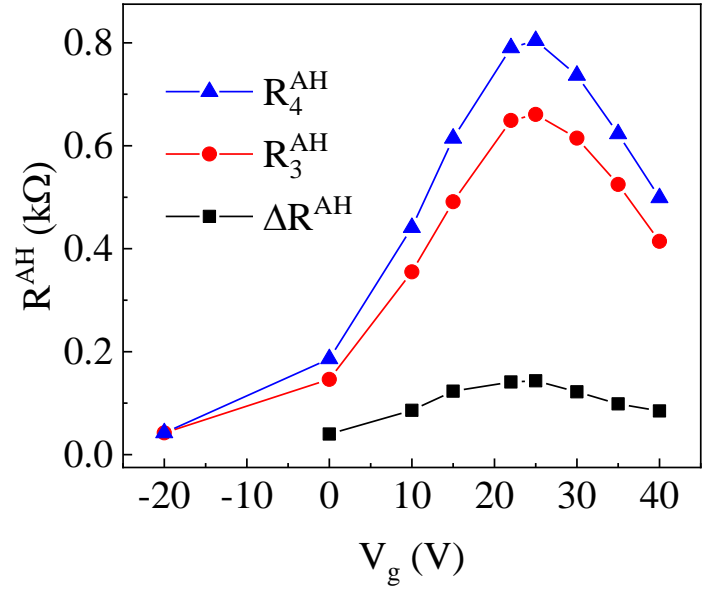

FIG. S3. The anomalous Hall resistance $R_4^{AH}$, $R_3^{AH}$ and the step height $\Delta R^{AH}$ in Device A as a function of back gate at 10 K.

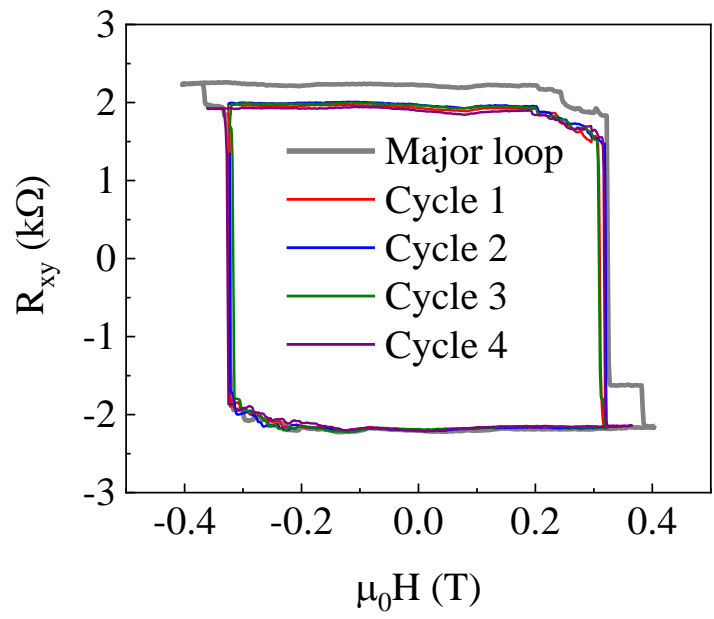

**FIG. S4. Repeated magnetic field cycling of the minor loop in Fig. 4(a) of main text.**

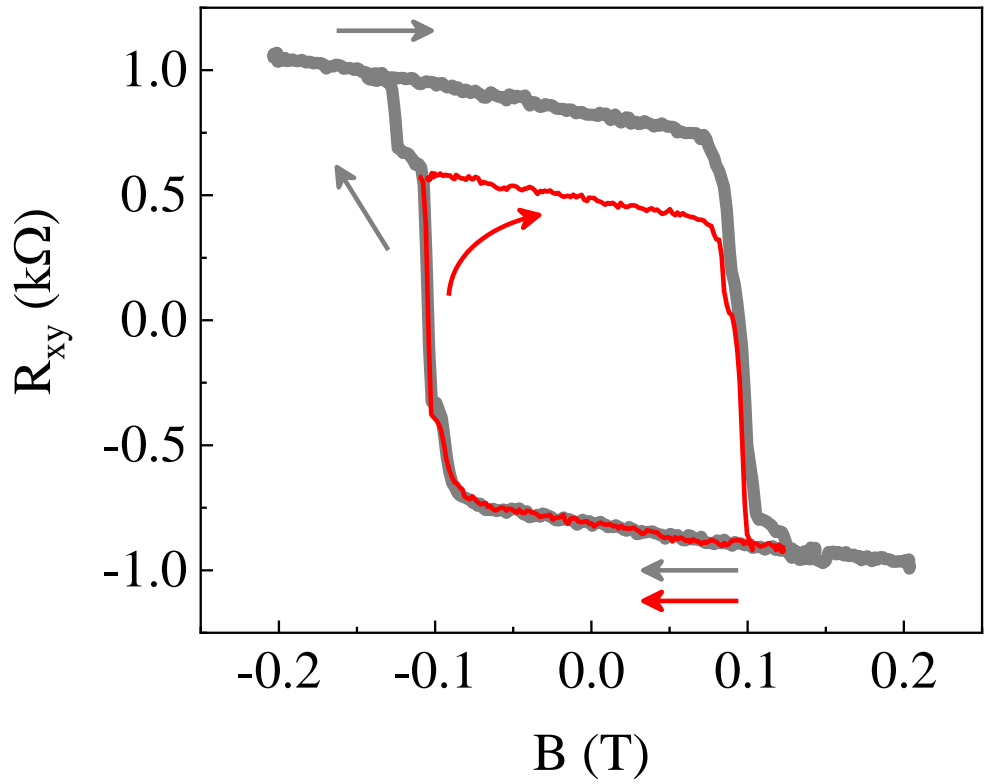

**FIG. S5. Multiple memory states in Device B at 15 K.** Major hysteresis loop (gray line) and minor loop (red line) were observed.

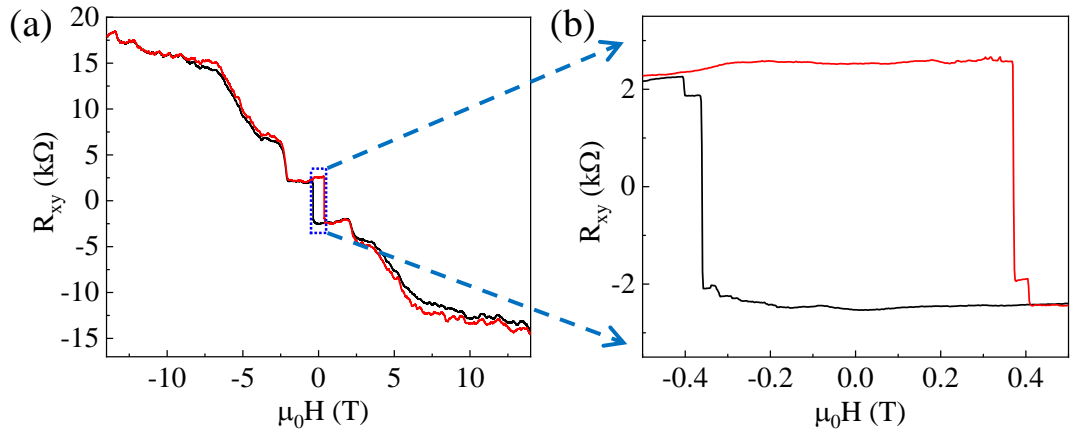

**FIG. S6. Robust intermediate Hall step behavior against large field cycles.**

(a) Four-terminal Hall resistance $R_{xy}$ as a function of magnetic field up to 14 T in Device A at 1 K and $V_g = 22$ V.

(b) Zoom-in view of (a). The double-step behavior of the hysteresis loop is robust against field cycle.

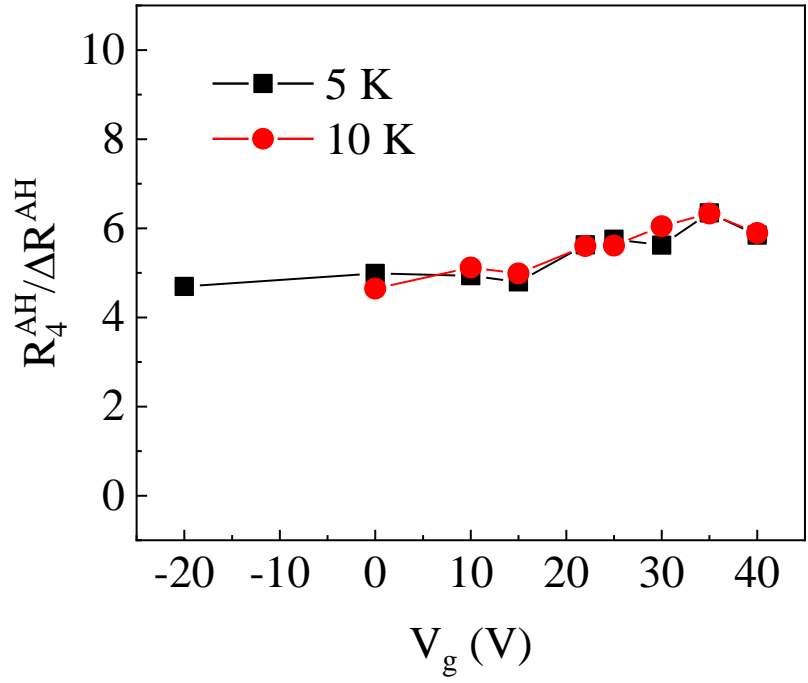

FIG. S7. The anomalous Hall ratio $R_4^{AH}/\Delta R^{AH}$ as a function of gate voltage at 5 K and 10 K.

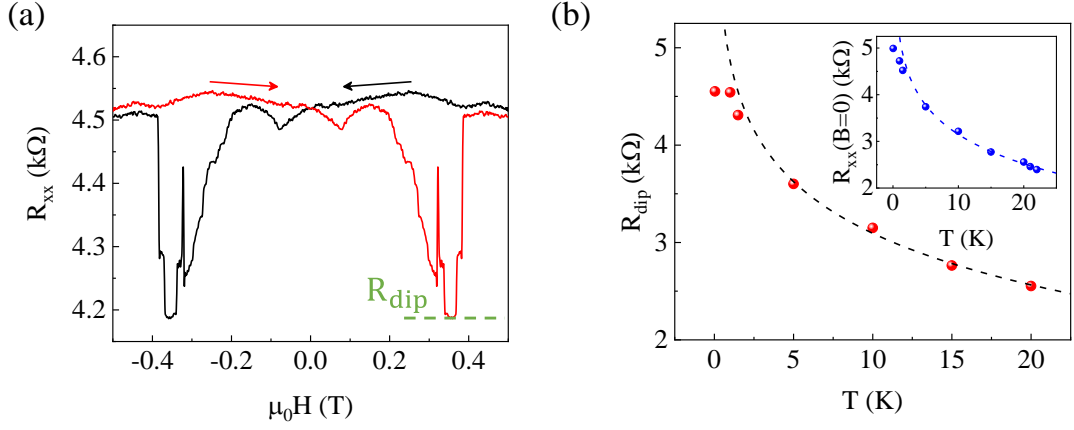

**FIG. S8. $R_{xx}$ dip and its temperature evolution.**

(a) Four-terminal longitudinal resistance $R_{xx}$ that corresponds to the major loop of $R_{xy}$ in Device A at 1.5 K and $V_g = 25$ V.

(b) Temperature evolution of $R_{dip}$. Inset: Temperature evolution of $R_{xx}(B=0)$. The dotted lines show $\ln(T)$-dependence. $R_{xx}(B=0)$ and $R_{dip}$ are defined as the resistance at zero field and the resistance dip, respectively.